\documentclass[12pt, a4paper, lenqo]{article}
\def\thetitle{\textbf{The Near Miss Effect \\ and the Framing of Lotteries}}
\usepackage[top=1in, bottom=1in,left=1in,right=1in]{geometry}
\usepackage[T1]{fontenc}
\usepackage{utopia}
\fontfamily{utopia}

\usepackage{amsmath,amssymb,mathtools}
\usepackage{bm}
\usepackage{longtable}
\usepackage{booktabs}

\DeclarePairedDelimiter\floor{\lfloor}{\rfloor}

\usepackage{setspace}

\usepackage[usenames,dvipsnames,svgnames,table]{xcolor}
\definecolor{CombinatoricaAqua}{HTML}{00698C}
\definecolor{CombinatoricaBlue}{HTML}{3A3293}
\definecolor{CombinatoricaBrown}{HTML}{66220C}
\definecolor{CombinatoricaRed}{HTML}{DF2A27}
\definecolor{HarvardCrimson}{rgb}{0.6471, 0.1098, 0.1882}

\usepackage[backref=page]{hyperref}
\hypersetup{
	unicode,
	pdfencoding=auto,
	pdfauthor={Michael Crystal},
	pdftitle={\thetitle},
	pdfsubject={},
	pdfkeywords={},
	colorlinks=true,
	citecolor=CombinatoricaBlue,
	linkcolor=CombinatoricaAqua,
	anchorcolor=CombinatoricaBrown,
	urlcolor=HarvardCrimson}

\usepackage{scalerel}

\usepackage{enumitem}

\usepackage{csquotes}

\usepackage{tikz}
\usepackage[justification=centering]{caption}
\usepackage{float}

\makeatletter
\let\reftagform@=\tagform@
\def\tagform@#1{\maketag@@@
	{(\ignorespaces\textcolor{Black}{#1}\unskip\@@italiccorr)}}
\renewcommand{\eqref}[1]{\textup{\reftagform@{\ref{#1}}}}
\makeatother

\usepackage{amsthm}
\usepackage{thmtools}
\usepackage{bbm}
\usepackage[capitalize]{cleveref}
\Crefname{fact}{Fact}{Facts}
\Crefname{claim}{Claim}{Claims}
\Crefname{assumption}{Assumption}{Assumptions}

\declaretheoremstyle[
spaceabove=\topsep, spacebelow=\topsep,
headfont=\color{CombinatoricaBlue}\normalfont\bfseries,
bodyfont=\itshape,
]{thm}
\declaretheoremstyle[
spaceabove=\topsep, spacebelow=\topsep,
headfont=\color{CombinatoricaBlue}\normalfont\bfseries,
bodyfont=\normalfont,
]{dfn}
\declaretheoremstyle[
spaceabove=0.5\topsep, spacebelow=0.5\topsep,
headfont=\color{CombinatoricaBlue}\normalfont\bfseries,
bodyfont=\normalfont,
]{rmk}

\declaretheorem[style=thm]{corollary}
\declaretheorem[style=thm]{proposition}

\declaretheorem[style=thm]{observation}

\declaretheorem[style=dfn]{definition}
\declaretheorem[style=dfn]{example}

\title{\thetitle
		\thanks{I am grateful to Rani Spiegler for his valuable guidance and comments. I thank Yair Antler, Ayala Arad, Itzhak Gilboa, Simon Litsyn, Paul Milgrom, Zvika Neeman, Ariel Rubinstein, and Itzhak Tamo for helpful discussions and comments. I also thank participants of various Stanford and Tel Aviv University workshops and seminars for comments and suggestions.}}

\author{Michael Crystal
	\thanks{Stanford University, Department of Economics; mcryst@stanford.edu}}

\date{}

\usepackage{apacite}
\usepackage{natbib}

\date{\today}

\begin{document}

\maketitle

\begin{abstract}
	We present a framework for analyzing the near miss effect in lotteries. A decision maker (DM) facing a lottery, falsely interprets losing outcomes that are close to winning ones, as a sign that success is within reach. As a result, the DM will prefer lotteries that induce a higher frequency of near misses, even if the underlying probability of winning is constant. We define a near miss index and analyze the optimal lottery design. This analysis leads us to establish a fruitful connection between our near miss framework and the field of coding theory. Building on this connection we compare different lottery frames and the near miss effect they induce. Analyzing a buyer-seller interaction allows us to gain further insight into the optimal framing of lotteries and might offer a potential explanation as to why lotteries with a very small probability of winning are commonplace and attractive.
	
\end{abstract}
	\textcolor{CombinatoricaBlue}{Keywords:} \textsl{Near Miss, Framing, Coding theory, Covering codes.}
\newpage
\section{Introduction}
	A miss is as good as a mile. This principle points to the fact that failing to achieve a certain goal is considered a failure, regardless of how close one may have come to success. 
	
	A rational decision maker (DM) should follow this principle in lotteries that are governed by pure chance. That is because coming close to success does not provide any useful information regarding the DM's ability or any indication that success is imminent. For instance, consider a lottery in which $n$ independent and identically distributed numbers are randomly drawn and participants receive a prize only if they guessed all numbers correctly. In this lottery, one should not hold different interpretations for guessing correctly $n-1$ numbers or none of them.
	
	Nevertheless, studies have shown that many individuals falsely interpret events in which they come very close to the desired outcome, also known as near miss events, as an indication of imminent success. \cite{Ski53} was the first to identify the near miss effect by claiming that in games of pure chance near miss events reinforce continued play. Later studies strengthened his claim by showing that individuals indeed react differently to different kinds of misses (see \cite{Cla09}, \cite{Fox12}, \cite{Sun12}, and \cite{Sta17} for such studies).\footnote{A different kind of near miss event that we do not analyze in the paper is a disaster that did not occur of was avoided (e.g. a car accident or a hurricane). \cite{Dil12} show that when people interpret these near miss events as disasters that did not occur, they illegitimately underestimate the danger of subsequent hazardous situations and make riskier decision.}
	
	\cite{Rei86} argued that near misses in slot machine pay-lines resulted in longer playing time. He also noted that some slot machine manufacturers are aware of that and take that into account in the design of the slot machines: 
	\begin{displayquote} 
	Commercial gambling systems, particularly instant lotteries and slot machines, are contrived to ensure a higher frequency of near misses than would be expected by chance alone. (p. 32)
	\end{displayquote}
	The Nevada Gaming Commission addressed this issue with a 1989 ruling that banned the design and programming of higher frequencies of near misses than chance alone.\footnote{Nevada Gaming Commission (1988, 1989). Nevada State Gaming Control Board v. Universal Distributing of Nevada. Case-No. 88-4.}
	
	This paper proposes an approach to modeling the near miss effect induced by different lottery frames and its influence on the subjective valuation of lotteries. To model the near miss effect we need to describe the relevant lottery framing and the metric used to measure the distance between any two possible outcomes. Motivated by the framing of slot machines, our lottery outcomes are characterized by vectors of length $n \in \mathbb{N}$\, in which each coordinate attains values from an alphabet of symbols $Q$ of size $q \in \mathbb{N}$. This possible set of outcomes, denoted by $Q^n$ along with a subset of winning outcomes $W\subseteq Q^n$\, fully describes a lottery in our setting. Therefore, a lottery is depicted by the triplet $\left(q,n,W\right)$. We assume that there is no inner metric in our alphabet of symbols and thus a natural measure for the distance between lottery outcomes is the number of coordinates in which they differ.
	
	We define a near miss index that captures the near miss effect induced by a given lottery and obtains values between zero and one. This index allows us to draw comparisons between different lotteries. Notably, we can form clear distinctions between different lotteries that have the same objective probability of winning, but entail a different framing that affect the frequency of near miss events. 
	
	Near misses might influence the DM's belief formation in the following sense: A rational DM taking part in a certain lottery will eventually form a correct belief over the probability of winning because he views all losing outcomes identically. However, for a non-rational DM, losing outcomes that are close to winning ones might provide positive feedback that will lead to a distortion in his probabilistic assessment and result in a higher perceived probability of winning. This distortion might stem from near miss event triggering the "close" counterfactual of a winning outcome (\cite{Kah90}, \cite{Kah82}, \cite{Lew73}).\footnote{Close counterfactuals are alternatives to reality that "almost happened". In our setting, a winning outcome might be a close counterfactual to losing outcomes that are "close" to it.} Thus, one possible interpretation of the near miss index is the DM's perceived probability of winning. Therefore, a different lottery framing might have a substantial impact on the subjective lottery valuation. This leads us to the following question: Given an objective probability of winning $p$, what is the lottery framing that maximizes the perceived probability of winning.\footnote{An alternative interpretation for the near miss effect is that some people derive additional utility from the thrill of being close to winning and enjoy gambling for its own sake (see \cite{Bin20}). This interpretation will also lead to higher subjective valuation of lotteries with higher near miss index.}
	
	The search for an optimal lottery frame takes us down an unexpected path to the field of information theory and connects our near miss index with coding theory and the study of covering codes (see \cite{Coh97} for a textbook). 
	
	We show that designing an optimal lottery is an open problem that generalizes the thoroughly studied \textsl{covering code problem} (\cite{Gol64}, \cite{Sch68}, \cite{Rod70}, \cite{Ost98}, \cite{Ker05}). Although the optimal lottery frame remains out of reach, we are still able to rank general lottery frames in terms of the near miss effect they generate. That is achieved by modifying results from coding theory and adapting them to our near miss analysis. We will show that by setting the lottery's parameters accordingly we obtain lottery frames that are preferable in terms of the near miss effect they induce. 
	
	The paper proceeds as follows: Section 2 introduces the near miss framework, describing the set of lotteries that will be analyzed and the near miss index definition for these lotteries. Also, the connection between the near miss index and the field of coding theory is established. Section 3 presents the optimal lottery frame analysis and compares different lottery frames and the near miss effect they induce. Furthermore, bounds on the near miss index are derived. Section 4 describes an interaction between a seller, who designs a lottery, and a potential buyer contemplating participation and payment for this lottery. We show that, in this setting, we can characterize the seller's optimal lottery frame. Section 5 provides concluding remarks.

\section{Near Miss Framework}

	Let $Q$ be a finite set of size $q \in \mathbb{N}$ that describes the \textsl{alphabet} of symbols, and let $n\in \mathbb{N}$ be the \textsl{length} of a possible lottery outcome. Denote by $Q^n$ the set of all possible lottery outcomes. Thus, each lottery outcome $x\in Q^n$ is represented by an n-dimensional vector in which each coordinate may obtain a finite set of values depicted by the alphabet. We denote by $W \subseteq Q^n$ the subset of winning outcomes. Formally, a lottery frame is defined as follows: 
	
	\begin{definition}[\textbf{Lottery Frame}]
		A lottery frame is defined by the triplet $\left(q,n,W\right)$, where $q$ is the alphabet size, $n$ is the length of a possible outcome, and  $W \subseteq Q^n$ is the set of winning outcomes. 	
	\end{definition}
	
	Our lottery space consists of all possible lotteries that have $q^{n}$ outcomes. For ease of notation and tractability we restrict attention to lotteries with a single prize and a uniform distribution over $Q^n$. Therefore, the probability of winning, denoted by $p$, is given by $p=\frac{|W|}{q^{n}}$. This restriction is without loss of generality and the analysis carried out and all the results that follows extend to general distributions over outcomes.

\subsection{Near Miss index}
	
	The definition of a near miss index must incorporate a clear distinction between different losing outcomes and their proximity to winning ones. For this distinction we need to define a measure of distance between any two possible outcomes in our lottery space. To address that, we will use the Hamming distance, which was introduced by \cite{Ham50}.
	
	\begin{definition}[\textbf{Hamming distance}]
		The Hamming distance between two equal-length vectors $\mathrm{\boldsymbol{x}}=\left(x_{1},x_{2},\ldots,x_{n}\right)$, $\mathrm{\boldsymbol{y}}=\left(y_{1},y_{2},\ldots,y_{n}\right)$ is the number of coordinates in which they differ, i.e.,
		
		\begin{equation}
		d\left(\mathrm{\boldsymbol{x}},\mathrm{\boldsymbol{y}}\right)= \hspace{0.1cm} \mid\left\{ i\colon x_{i}\neq y_{i}\right\} \mid \\[5pt]
		\end{equation}
		
	\end{definition}

	In employing the Hamming distance, we implicitly make the assumption that the distance between any two lottery outcomes depends solely on the number of coordinates in which they differ and not on the specific values these coordinates attain. This assumption is plausible if the alphabet is comprised of unrelated symbols. For example, in a slot machine, it is not clear whether the symbol 7 is closer to a bar or a cherry. 
	However, if the alphabet is comprised of numbers this assumption seems less plausible since in that case, the alphabet has a clear, well defined inner metric.\footnote{For instance, guessing that 6 is one of the winning numbers in a lottery ticket or guessing that 2 is, will likely be viewed differently by the DM if the winning number turns out to be 7. For these settings, one might think of different distance measures that incorporate an inner metric of the alphabet (e.g. numbers).} Nevertheless, even in scenarios in which the alphabet has a well defined inner metric, our index can be perceived as capturing a first-order effect. Specifically, this is true when individuals prioritize the quantity of correct guesses, placing greater emphasis on this aspect before examining the distance between specific incorrect ones.
	
	Using the Hamming distance, we define the near miss index as follows.
	
	\begin{definition}[\textbf{Near Miss Index}]
		The Near Miss Index for a lottery $\left(q,n,W\right)$ is given by:
		
		\begin{equation}
			\mathcal{NM}\left(q,n,W\right) = \mathbb{E}_{x\in{Q^n}}\left[1-\frac{\min_{w\in{W}}d(x,w)}{n}\right]
		\end{equation}
		\\
	\end{definition}

	The near miss index is a weakly decreasing transformation of the expected minimal Hamming distance to a winning outcome, where $\min_{w\in W}d \left(x,w\right) $ is the minimal number of shifts needed to obtain a winning outcome given a certain outcome $x \in Q^n$. Note that for any given lottery the near miss index obtains values from zero to one, i.e. $ 0 \leq \mathcal{NM}\left(q,n,W\right) \leq 1 $. The division by the lottery's length takes into account that being $k$ shifts away from a winning outcome is intuitively closer to it as the lottery's length increases.
	
	As noted, a possible interpretation of the near miss index is the perceived probability of winning. Following this interpretation, a DM facing a lottery with a higher near miss index will form a belief that the underlying probability of winning is higher than the actual objective one. 
	
	\begin{example}[\textbf{Slot Machine}] ~\\
		Consider a slot machine comprised of three reels and two equally likely symbols - 7 and bar. 
		
		First, assume that there are two possible winning outcomes, either all three reels show 7 or all three show bar. The set of outcomes of this lottery is given by $\left\{7,B\right\}^{3}$. Each outcome is represented by a three dimensional vector in which each coordinate is either 7 or bar, i.e. $x=\left(x_{1},x_{2},x_{3}\right)$ where $x_{i} \in \left\{7,B\right\}$. Thus, this lottery can be represented by $\left(q,n,W\right)=\left(2,3,\left\{\left(7,7,7\right),\left(B,B,B\right)  \right\} \right)$ The near miss index for this lottery is given by:	
		
		\begin{equation*}
			\begin{split}
				\mathcal{NM}\left(2,3,\left\{\left(7,7,7\right),\left(B,B,B\right)  \right\} \right) 
				& = \frac{1}{4} \cdot \left(1-\frac{0}{3}\right) + \frac{3}{4} \cdot \left(1-\frac{1}{3}\right) \\[10pt]  
				& = \frac{1}{4}+\frac{3}{4}\cdot\frac{2}{3} = \frac{3}{4} \\[10pt]
			\end{split}	
		\end{equation*}
	
	Alternatively, take the winning set of outcomes to be $W^{'}=\left\lbrace \left(7,7,7\right), \left(B,B,7\right) \right\rbrace$. This lottery which is represented by $\left(q,n,W\right)=\left(2,3,\left\{\left(7,7,7\right),\left(B,B,7\right)  \right\} \right)$ obtains a near miss index of:
	
		\begin{equation*}
		\begin{split}
			\mathcal{NM}\left(2,3,\left\{\left(7,7,7\right),\left(B,B,7\right)\right\}\right)
			& = \frac{1}{4} \cdot \left(1-\frac{0}{3}\right) + \frac{2}{4} \cdot \left(1-\frac{1}{3}\right) + \frac{1}{4} \cdot \left(1-\frac{2}{3}\right) \\[10pt]  
			& = \frac{1}{4}+\frac{2}{4}\cdot\frac{2}{3} +\frac{1}{4}\cdot\frac{1}{3} = \frac{2}{3} \\[10pt]
		\end{split}	
		\end{equation*}	
		
	Note that both lotteries have the same objective probability of winning, however, they differ in their near miss index and thus might induce different perceived winning probability. We can see that even if we keep $q$, $n$ and the size of $W$ fixed, changing the specific winning outcomes might result in a different near miss index.
	
	Moreover, we can see that when $W=\left\lbrace \left(7,7,7\right), \left(B,B,B\right) \right\rbrace$, the Hamming distance from any losing outcome to a winning one is at most one. Therefore, if we set the objective probability of winning at $p=\frac{1}{4}$, this specification of $W$ along with $q=2$ and $n=3$  maximizes the near miss index.
	\hfill $\diamondsuit$ 
	\end{example}
	
	A natural question that arises is: Can we increase the perceived probability of winning by changing the triplet $\left(q,n,W\right)$ while keeping the objective probability of winning constant at some $p$.
	 
\subsection{Near Miss and Codes}

	  Coding theory focuses on the study of codes and their properties. It is relevant to several scientific fields such as information theory and computer science. We present a concise description of covering codes and the connection to our near miss framework.
	  
	  \begin{definition}[\textbf{$q$-ary codes}]
	  	\textsl{Let $Q$ be a finite set with q elements. A nonempty subset $W$ of $Q^{n}$ is called a $q$-ary code of length n.}
	  \end{definition} 
  
	  For instance, in Example 1 the alphabet is given by $Q=\left\{7,B\right\}$, which is the binary case. The set $Q^{n}$ is called the $n$-dimensional $q$-ary Hamming space. The vectors belonging to a code $W$ are called \textsl{code-words}. Thus, the winning set of outcomes corresponds to a $q$-ary code where each winning outcome corresponds to a code-word. 
	  
	  \begin{definition}[\textbf{Covering Radius}]
	  	\textsl{
	  		The covering radius of a code $W\subseteq Q^n $ is the smallest integer $R$ such that every vector $ x\in Q^n$ is $R$-covered by at least one code-word of \hspace{0.1mm} $W$, i.e.,}
	  		\begin{equation}
	  		R =  \max_{x\in Q^n}\min_{w\in W}d\left(x,w\right) 
	  		\end{equation}
	  		\textsl{Where $d\left(x,w\right)$ is the hamming distance.}
	  \end{definition}
  
  	  In other words, the covering radius of a code $W$ is the distance between the code and the farthest-off vectors (i.e. code-words) in the Hamming space. A code $W$ with a covering radius $R$ is called a $q$-ary $R$-covering code.
  	  
  	  Denote by $K_{q}\left(n,R\right)$ the minimal cardinality of a $q$-ary $R$-covering code with length $n$. That is, $K_{q}\left(n,R\right)$ is the minimal number of code-words needed to ensure that every code-word in the $n$-dimensional $q$-ary Hamming space is $R$-covered. The following table presents some known values of $K_{q}\left(n,R\right)$ for the binary case and $R\in\left\lbrace 1,2 \right\rbrace $.\footnote{See \url{http://old.sztaki.hu/~keri/codes/2_tables.pdf} for known values of $K_{2}\left(n,R\right)$.} \\[3pt]
  	  
  	  \begin{table}[H]
  	  	\fontsize{11}{9}
  	  	\centering
  	  	\begin{tabular}{c c c c c c c c c c c}
  	  		\toprule
  	  		\textbf{$K_{2}(n,R)$} & $n=1$ & $n=2$ & $n=3$ & $n=4$ & $n=5$ & $n=6$ & $n=7$ & $n=8$ & $n=9$ & $n=10$   \\
  	  		\midrule
  	  		\midrule
  	  		$R=1$ & 1 & 2 & 2 & 4 & 7 & 12 & 16 & 32 & 62 & 107-120 \\
  	  		\midrule
	  	  		$R=2$ & & 1 & 2 & 2 & 2 & 4 & 7 & 12 & 16 & 24-30 \\
  	  		\bottomrule
  	  	\end{tabular}
    	\captionsetup{justification=centering}
  	  	\caption{$K_{2}\left(n,R\right)$ - Size of minimal binary covering codes} 
  	  \end{table}
    
    We can see from Table 1 that even for the relatively simple case of $q=2$, if we set $n=10$ the exact value of $K_{q}\left(n,R\right)$\, is unknown. Finding $K_{q}\left(n,R\right)$\, for finite parameters $q,n$ and $R$\, is called the \textsl{covering code problem}. This problem is an open one and there is no known algorithm or constructive solution for it. 
    
    Note that an optimal construction of minimal covering codes yields an optimal lottery frame for certain winning probabilities. For example, assume that the objective probability of winning is given by $p=\frac{12}{2^6}=\frac{3}{16}$, taking a lottery frame with alphabet size $q=2$, length $n=6$, and constructing the set winning outcomes $W$ to be the minimal 2-ary 1-covering code of length $n=6$ will result in an optimal lottery for this objective probability. As shown in Table 1 that will require 12 distinct winning outcomes.
    Thus, finding an optimal near miss index for any given objective probability of winning $p$ generalizes the covering code problem.
    
    Although a construction of covering codes remains unknown, adapting several results from coding theory and modifying them to fit our near miss analysis will allow us to compare general lottery frames.
    
\section{Lottery Frames} 

	In this section we will analyze different lottery frames, keeping the objective probability of winning fixed. We will show that certain frames induce a higher near miss effect. 

	We start by examining the case where the number of possible lottery outcomes is constant and allow for variation in $n$ and $q$. We show we can always weakly increase the near miss index of a lottery frame by increasing the length of the lottery while reducing the alphabet size. 
	
	\begin{proposition}
		Fix an objective probability of winning $p$. For any  $t\in\mathbb{N}$ and any lottery $\left(q^{t},n,W\right)$ there exist a lottery $\left(q,tn,W^{'}\right)$ such that:
		
		\begin{equation}
			\mathcal{NM}\left(q,tn,W^{'}\right) \: \geq \: \mathcal{NM}\left(q^{t},n,W\right)
		\end{equation}
		
	\end{proposition}

	\noindent
	\textbf{Proof.} Given a lottery $\left(q^{t},n,W\right)$ we will construct the lottery $\left(q,tn,W^{'}\right)$ such that $|W|=|W^{'}|$ and (3) holds.
	
	Let $Q$ be the alphabet of size $q$ and define the product set $Q\times \dots \times Q$ to be the alphabet of size $q^{t}$.
	Denote by $g: \left(Q^{t}\right)^{n}  \to Q^{tn}$ the bijective mapping that takes a vector of length $n$ over the alphabet $Q\times \dots \times Q$ and maps it to a vector of length $tn$ over the alphabet $Q$ by expanding each of the $n$ coordinates to a vector of length $t$. For example, let $Q=\left\{7,B\right\}$, $t=2$, $n=2$, and $x = \left(7B,77\right) \in \left(Q\times Q\right)^{2}$. Expanding each coordinate of $x$ to a vector of length 2 we obtain $g(x)=\left(7,B,7,7\right) \in Q^{4}$. 
	
	Given a winning set of outcomes $W$ which is defined by vectors of length $n$ over the alphabet $Q\times \dots \times Q$  we will construct the winning set of outcomes $W^{'}$ by $g\left(W\right)$. 
	
	Note that $|W|=|W^{'}|$ is immediate from construction. Moreover, for each possible outcome $x\in \left(Q^{t}\right)^{n}$ the following holds:
	
	\begin{equation*}
		t\cdot\left(\min_{w\in W}d\left(x,w\right)\right) 
		\geq
		\min_{g\left(w\right)\in g\left(W\right)}d\left(g\left(x\right),g\left(w\right)\right)
	\end{equation*}
	
	The above inequality holds due to the following: Assume that for a given outcome $x \in \left(Q^{t}\right)^{n}$ the following $\min_{w\in W}d\left(x,w\right)=k$ holds for some $k \in \mathbb{N}$. Take any coordinate in which $x$ and $w$ differ. Expanding this coordinate in both $x$ and $w$ to vectors of length $t$ will result in two vectors that differ by at most $t$ coordinates. Thus, the Hamming distance between $g(x)$ and $g(w)$ will be at most $t\cdot k$. 
	
	Therefore,
	
	\begin{equation*}
		\begin{split}
			\mathcal{NM}\left(q,tn,W^{'}\right) 
			&= 
			1-\frac{1}{tn}\mathbb{E}_{g\left(x\right)\in  g\left( \left(Q^{t}\right)^{n}\right) }\left[\min_{g\left(w\right)\in g\left(W\right)}d\left(g\left(x\right),g\left(w\right)\right)\right] \\[10pt] 
			&\geq
			1-\frac{1}{tn}\mathbb{E}_{x\in \left(Q^{t}\right)^{n}}\left[t\cdot\min_{w\in W}d\left(x,w\right)\right] \\[10pt] 
			&=
			1-\frac{1}{n}\mathbb{E}_{x\in \left(Q^{t}\right)^{n}}\left[\min_{w\in W}d\left(x,w\right)\right] \\[10pt] 
			&=
			\mathcal{NM}\left(q^{t},n,W\right)
		\end{split}
	\end{equation*} 
	\hfill $\qedsymbol$ \\[0.5pt] \par
	This result indicates that if our goal is to increase the near miss index and we were to design a slot machine we would increase the number of reels and reduce the number of symbols. However, there are feasibility constraints in designing a slot machine with many reels, one of them is the physical space it requires. Also, increasing the number of reels might complicate the DM's inference as to whether a given outcome is close or far from a winning one. This leads us to our next result.
	
	Assuming that the lottery's length is set at some $n \in \mathbb{N}$, increasing the alphabet size is weakly preferable in terms of near miss.  
	
	\begin{proposition}
		\textsl{Fix an objective probability of winning $p$. For any $t\in\mathbb{N}$ and any lottery $\left(q,n,W\right)$ there exists a lottery $\left(tq,n,W^{'}\right)$ such that:} 
		
		\begin{equation}
		\mathcal{NM}\left(tq,n,W^{'}\right) \: \geq \: \mathcal{NM}\left(q,n,W\right)
		\end{equation}
		
	\end{proposition} 
	\noindent
	\textbf{Proof.} Given a lottery $\left(q,n,W\right)$ we will construct the winning set of outcomes $W^{'}$ such that $|W^{'}|=|W| \cdot t^{n}$ and we will show that (4) holds.
	
	Let $Q=\{1,2,\dots,q\}$ be the alphabet of size $q$. Denote by $T=\{1_{1},1_{2},\dots,1_{t},2_{1},2_{_2},\\\dots,2_{t},\dots,q_{1},\dots,q_{t}\}$ the alphabet of size $tq$. We will construct the winning set of outcomes $W^{'}$ by associating with each vector $w=(a,b,\dots,c,d)$ in $W$ the following $t^n$ vectors:
	
	\begin{equation*}
		(a_{i},b_{j},\dots,c_{k},d_{l}), \; 1 \leq i,j,\dots,k,l \leq t 
	\end{equation*}
	
	Note that $|W^{'}|=|W|\cdot t^{n}$ is immediate from construction. Let $u=(u_{1},\dots,u_{n})$ be a vector in $Q^n$ and assume that $w\in{W}$ is the closest winning outcome to $u$. Denote by $U$ the $t^n$ vectors associated with $u$ and by $\bar{W}$ the $t^n$ vectors associated with $w$. Note that by construction, for each vector in $U$ there exists a vector in $\bar{W}$ located at a Hamming distance of $d(u,w)$. Therefore we have found a specific construction in which equation (4) holds with equality. Moreover, by setting $q=n=t=2$ and $p=\frac{1}{4}$ equation (4) holds with strict inequality. That is because these parameters imply that $|W|=1$ and $|W^{'}|=4$. This along with the fact that $K_{4}\left(2,1\right)=4$ and $K_{2}\left(2,1\right)=2$ completes the proof.\hfill $\qedsymbol$ \\[0.5pt]
	 
	Some real-life evidence from the design of slot machines over the years might suggest that these considerations were taken into account. From their creation, at the end of the nineteenth century to this day, the most common design is the three reel slot machine (i.e. the length of the lottery is $n=3$). However, the number of symbols in most slot machines increased from just five to around twenty (i.e. $q=20$). Moreover, some virtual slot machines have as many as 256 symbols.
	
	Moving forward, our next result shows that if we fix the lottery's alphabet size at some $q\in \mathbb{N}$, increasing the length will result in a higher near miss index. 
	
	\begin{proposition}
		
		Fix an objective probability of winning $p$. For any lottery $\left(q,n,W\right)$ there exists a lottery $\left(q,n+1,W^{'}\right)$ such that:
		
		\begin{equation}
			\mathcal{NM}\left(q,n+1,W^{'}\right) \: > \: \mathcal{NM}\left(q,n,W\right)\
		\end{equation}
	\end{proposition}
	\noindent
	\textbf{Proof.} Given a lottery $\left(q,n,W\right)$ we will construct the lottery $\left(q,n+1,W^{'}\right)$ such that $|W^{'}|=|W| \cdot q$ and (5) holds. 
	
	Let $Q$ denote the alphabet of size $q$. Given a lottery $\left(q,n,W\right)$ we will use $W$ and $Q$ to construct the winning set of outcomes $W^{'}$ by the following:
	
	\begin{equation*}
		W^{'}=W \oplus Q=\left\{ \left(w,w_{n+1}\right):w\in W,w_{n+1}\in Q\right\} 
	\end{equation*} 
	
	In other words, $W^{'}$ is the concatenation of all $n$-dimensional code-words $w\in W$ with all $1$-dimensional code-words $w_{n+1}\in Q$. 
	
	$|W^{'}|=|W|\cdot q$ \hspace{0.1mm} is immediate from construction. Note that the set of outcomes $Q^{n+1}$ can be written as $Q^n\oplus Q$. Thus, the minimal distance between an outcome $x^{'}=\left(x,x_{n+1}\right)\in Q^{n+1}$ to a winning outcome is given by:
	
	\begin{equation*}
		\min_{w^{'}\in W^{'}}d\left(x^{'},w^{'}\right)=\min_{w\in W}d\left(x,w\right)+\min_{w_{n+1}\in Q}d\left(x_{n+1},w_{n+1}\right)
	\end{equation*}
	
	Therefore $\min_{w_{n+1}\in Q}d\left(x_{n+1},w_{n+1}\right)=0$, which implies that $\min_{w^{'}\in W^{'}}d\left(x^{'},w^{'}\right)=\min_{w\in W}d\left(x,w\right)$. Thus, \hspace{0.1mm} $\mathcal{NM}\left(q,n+1,W^{'}\right) \: > \: \mathcal{NM}\left(q,n,W\right)$. The strict inequality holds due to the division by the lottery's length in the near miss index. \hfill $\qedsymbol$ \\[0.5pt]
	
	Therefore, keeping the number of symbols fixed we would benefit from increasing the number of reels in a slot machine. This result indicates that the physical constraint over the number of reels might be binding.  
	
	We have shown that if one parameter of the lottery's frame is held fixed increasing the other one is preferable in terms of near miss. Furthermore, if we allow for variation in both parameters, reducing the alphabet size while increasing the length will result in a preferable lottery frame.
	
\subsection{Near Miss Bounds}

	In this subsection, we move from qualitative to quantitative results and derive bounds on the ratio between the maximal near miss index and the lottery's objective probability of winning for a given lottery frame. This ratio is of interest because it describes the degree to which the objective probability of winning might be inflated. In what follows, we use the connection between our near miss framework and coding theory.
	
	\begin{proposition}
		\noindent
		\textsl{Let $n$ and $q$ be the length and alphabet size of the lottery respectively.} \\ 
		\begin{enumerate}[label=(\roman*)]
			\item Assume that $p \leq \frac{q^{2}+1}{2q^{3}}$. Then there exists a set of outcomes $W$ such that the following holds:
			
			\begin{equation}
				\frac{2q^{3}(n-1)+(q^{2}+1)}{n(q^{2}+1)}
				\: \leq \: 
				\frac{\mathcal{NM}\left(q,n,W\right)}{p} 
				\: \leq \:
				q^{n-1} 
			\end{equation}
			
			\vspace{0.25in}
			\item Assume that $p \geq \frac{q^{2}+1}{2q^{3}}$. Then there exists a set of outcomes $W$ such that the following holds:
			
			\begin{equation}
			1
			\: \leq \:
			\frac{\mathcal{NM}\left(q,n,W\right)}{p} 
			\: \leq \:
			\frac{2q^{3}(n-1)+(q^{2}+1)}{n(q^{2}+1)}
			\end{equation}
			
		\end{enumerate}
	\end{proposition}  
	
	\vspace{0.25cm}
	\noindent
	\textbf{Proof.} The proof proceeds by a series of steps. \\
	
	\noindent \textbf{Step 1: Concave near miss.} \\
	We begin by showing that the maximal near miss index as a function of the objective probability of winning is concave. To show that we will prove an equivalent claim, that the minimal sum of distances $\sum_{x\in X_{q}^{n}}\min_{w\in W}d(x,w)$ is convex in $p$.
	Assume by contradiction that this sum is not convex in $p$ and let $p_{m}=\frac{m}{q^{n}}$ be the smallest probability which violates convexity.
	
%
%
%
%
%

	Denote by $d_{1}^{*}$ and $d_{m}^{*}$ the minimal sum of distances for $p_{1}=\frac{1}{q^n}$ and $p_{m}$ respectively and let $W_{m}$ be the set of winning outcomes of size $m$ achieving $d_{m}^{*}$. We can reorder $W_{m}$ in a greedy manner such that in each step the added winning outcome results in the greatest reduction in the sum of distances. Let \,$W_m^G$\, be the reordered set, the path of \,$W_{m}^G$\, from $d_{1}^{*}$ to $d_{m}^{*}$ is convex and thus must pass strictly below the path of \,$W_m$\, for \,$p_{m-1}$. This contradicts the minimality of the distances sum. \hfill $\qedsymbol$  \\
	
	\noindent \textbf{Step 2: $\textbf{M}$-Bound.}\\
	Denote by $\mathcal{NM}^{*}$ the maximal near miss index for a given objective probability of winning $p^{*}$ and let $M=\frac{\mathcal{NM}^{*}}{p^{*}}$. By the concavity of the maximal near miss index function it follows that for every $p\leq p^{*}$ there exists a set of winning outcomes such that the following holds:
	
	\begin{equation*}
		M 
		\leq 
		\frac{\mathcal{NM}\left(q,n,W\right)}{p} 
	\end{equation*} \\
	\noindent Furthermore, for every $p\geq p^{*}$ the following holds: \\
	\begin{equation*}
		\frac{\mathcal{NM}\left(q,n,W\right)}{p}
		\leq
		M 
	\end{equation*}	
	\hfill $\qedsymbol$
	
	\noindent \textbf{Step 3: Showing that}\\
	For $p=\frac{q^{2}+1}{2q^{3}}$, the $M$-bound is given by: 
	\begin{equation}
	M=\frac{2q^3(n-1)+q^{2}+1}{n(q^{2}+1)}
	\end{equation}
	
	\noindent We will integrate two known results from coding theory to show that (8) holds. \\
	
	\noindent \textbf{Theorem 1 (\cite{Kal69}).} $K_{q}(3,1)=\floor{\frac{q^{2}+1}{2}}$. \\
	
	\noindent \textbf{Theorem 2 (\cite{Coh97}).}\footnote{This Theorem is a well known folklore result in coding theory. See \cite{Coh97} page 63.} $K_{q}(n,R)\leq qK_{q}(n-1,R)$. \\
	\noindent
	Integrating these results we obtain:
		\begin{equation*}
			\begin{split}	
				K_{q}(n,1) & \leq q^{n-3}K_{q}(3,1) \\[5pt]
				& \leq \frac{q^{n-3}(q^{2}+1)}{2}
			\end{split}
		\end{equation*}
		
		Therefore, if $|W|=\frac{q^{n-3}(q^{2}+1)}{2}$ we can construct a set of winning outcomes such that each losing outcome is located within a Hamming distance of one from a winning one. Thus, the $M$-bound is given by:
			\begin{equation*}
				\begin{split}
					M & =\frac{\mathcal{NM}\left(q,n,W\right)}{p} \\[5pt]
					& =\frac{2q^{3}(n-1)+q^{2}+1}{n(q^{2}+1)}
				\end{split}
			\end{equation*}
		\hfill $\qedsymbol$
	
	\noindent \textbf{Step 4: The remaining bounds}\\
	We will complete the proof by showing that the upper bound in (6) and the lower bound in (7) holds.
	
	To prove that the upper bound in (6) holds we will look at $|W|=1$ which implies $p=\frac{1}{q^{n}}$. Given only one winning outcome, the average Hamming distance of the losing outcomes to the winning one is: \\[5pt]
		\begin{equation*}
			\frac{\sum_{k=1}^{n}\binom{n}{k}k(q-1)^{k} }{q^{n}}=\frac{n(q-1)q^{n-1}}{q^n}=\frac{n(q-1)}{q}
		\end{equation*} \\
		
	Therefore, the near miss index for $p=\frac{1}{q^n}$ is $\frac{1}{q}$ and the $M$-bound is given by $M=q^{n-1}$.

	Finally, by the definition of the near miss index $p\leq \mathcal{NM}\left(q,n,W\right)$ holds for every $p$. Note that for $p=1$ this lower bound holds with equality. \hfill $\qedsymbol$ \\
	
	Note that the $M$-bound is increasing in both $q$ and $n$. Therefore, by increasing these parameters we attain an increased magnitude of the near miss effect relative to the objective probability of winning. To get a quantitative sense of Proposition 4, consider the following table, which provides values of the $M$-bound and the threshold objective probability of winning $p$ for various specifications of $q$ and $n$:
	
	\begin{table}[H]
		\fontsize{11}{9}
		\centering
		\begin{tabular}{c c c c}
			$q$ & $n$ & $p(q)$ & $M$ \\
			\midrule
		
			2 & 2 & 0.3125 & 2.10 \\
			
			2 & 10 & 0.3125 & 2.98\\
			
			5 & 2 & 0.1040 & 5.31\\
			
			5 & 10 & 0.1040 & 8.75\\
			
			8 & 2 & 0.0635 & 8.38\\
			
			8 & 10 & 0.0635 & 14.28\\
			
			11 & 2 & 0.0458 & 11.41\\
			
			11 & 10 & 0.0458 & 19.74\\
			
			14 & 2 & 0.0359 & 14.43\\
			
			14 & 10 & 0.0359 & 25.17\\
			
		\end{tabular}
	\end{table}
	 
	This table illustrates how $q$ and $n$ affect the near miss index achieved. For example, if we set $q=2$ and $n=2$, then for any objective probability of winning $p \leq 0.3125$ we can obtain a near miss effect of $2.1p$ or more. However, if we choose the objective probability of winning to be $p \geq 0.3125$ we can obtain a near miss effect of at most $2.1p$. 
	
\section{Selling a Framed Lottery}

	In this section, we present an interaction between a seller who designs a lottery and a buyer (DM) who considers buying it.
	
	The DM who faces the lottery has a misspecified belief that the probability of winning is the lottery's near miss index. Thus, assuming that the DM's willingness to pay is depicted by the near miss index and the seller's cost is given by the objective probability of winning, the seller wishes to design a lottery that maximizes the difference between them. We will refer to this difference as the seller's value. Therefore, the seller maximizes the following: \vspace{0.1cm}
	
	\begin{equation}
		\max_{\left(q,n,W\right)}\left[\mathbb{E}_{x\in{Q^n}} \left[ 1-\frac{\min_{w\in{W}}d(x,w)}{n}\right]-p\right]
	\end{equation}
	\\
	\indent As we have seen, maximizing the near miss index is an open problem. However, in this setting, we will show that if the seller chooses an objective probability of winning from a certain range, the optimal lottery frame can be fully characterized by parameters identical to those of a special family of codes called Hamming codes. We start by stating the following upper bound on the seller's value. \vspace{0.1cm}
	
	\begin{observation}
		Fix the probability of winning at some $p^{*}$. Then for any lottery frame $\left(q,n,W\right)$ with an objective probability of winning $p = \frac{|W|}{q^n}  \geq p^{*} $the following holds:
		
		\begin{equation}
		\max_{\left(q,n,W\right)}\left[\mathbb{E}_{x\in{Q^n}} \left[ 1-\frac{\min_{w\in{W}}d(x,w)}{n}\right]-p\right] 
		\: \leq \:
		\frac{n-1}{n}\cdot(1-p^{*})
		\end{equation}
		\vspace{0.1cm}
	\end{observation}	

	In other words, the seller's value is bounded above by setting $p=p^{*}$ and constructing a lottery in which every losing outcome is located within the minimal Hamming distance possible of exactly one from a certain winning outcome. 
		
	A code $W \subseteq Q^n$ with covering radius $1$ is called a Hamming code if for every code-word $x \in Q^n \setminus W$ there is a unique code-word $w \in W$ that is located within a Hamming distance of one from it. 
	Hamming codes were introduced by \cite{Ham50} and a full characterization of their parameters is known.\footnote{See \cite{Coh97} chapters 2 and 11 for a more detailed description of Hamming codes that includes a general algorithm for generating such codes.} They are defined by the following parameters: $q$ is a prime power, $n=\frac{q^{m}-1}{q-1}$ where $m\geq1$ and $R=1$. For these codes it is known that $K_{q}\left(n,1\right)=q^{n-m}$.

	Hamming codes are used in the field of coding theory for data compression and as error-correcting codes due to their efficiency.\footnote{Error correcting codes are codes used to overcome possible errors in data transmission caused by an unreliable or noisy communication channel.} In our framework this efficiency will be translated into maximizing the difference between the near miss index and the objective probability of winning.
	
	Our next result connects Hamming codes with the seller's maximization problem, showing that the optimal lottery frame has the same parameters.
 	
 	\begin{proposition}
 	    Assume that $p\geq\frac{1}{q^{m}}$ where $q$ is a prime power and $m\geq1$. Then the optimal lottery frame is defined by the following parameters: $q$, $n=\frac{q^m-1}{q-1}$ and $|W|= q^{n-m}$. 
 	\end{proposition}
	
	\noindent
	\textbf{Proof.} We start by setting the length of the lottery to be $n=\frac{q^{m}-1}{q-1}$ and the size of $W$ to $q^{n-m}$. By doing that, we obtain a lottery with an objective probability of winning $p=\frac{|W|}{q^{n}}=\frac{1}{q^{m}}$ that has the same parameters as those of a Hamming code. As noted, for these codes the following holds $K_{q}\left(n,1\right)=q^{n-m}=|W|$. Thus, there exists a set of outcomes $W$ of size $q^{n-m}$ such that for any $x \in Q^n \setminus W$ the following $\min_{w\in W}d\left(x,w\right) = 1$ holds. Therefore, the seller's value obtained from this lottery is given by: \\
	\begin{equation*}
		\begin{split}
		\mathbb{E}_{x\in Q^n} \left[ 1-\frac{\min_{w\in{W}}d(x,w)}{n}\right]-p 
		&= 1-\frac{0}{n}\cdot p - \frac{1}{n} \cdot (1-p) -p \\[5pt]
		&= \frac{n-1}{n} \cdot (1-p) \\[5pt]
		&= \frac{n-1}{n} \cdot \frac{q^m-1}{q^m}
		\end{split}
	\end{equation*}
	\\
	\noindent Thus, we have attained the upper bound on the seller's value for $p\geq\frac{1}{q^{m}}$ which concludes the proof. \hfill $\qedsymbol$ \\[0.25pt] \par
	 
	Therefore, by restricting attention to winning probabilities depicted by $p \geq \frac{1}{q^m}$ where $m\geq1$ and selecting an alphabet size that is a prime power, we can obtain a closed solution for the seller's optimal lottery frame.
	
	The following result shows that given a probability of $p=\frac{1}{q^{m}}$, the minimal length needed to obtain the seller's maximal value is that of a Hamming code.
	
	\begin{proposition}
		Assume that $p=\frac{1}{q^{m}}$ where $q$ is a prime power and $m\geq1$. Then the minimal length needed to obtain the seller's maximal value is $n=\frac{q^{m}-1}{q-1}$.
	\end{proposition}

	\noindent \textbf{Proof.} To prove this result we will use a lower bound of covering codes called the sphere covering lower bound which is only attained by perfect codes.\footnote{Hamming codes are a family of perfect codes.} \\
	
	\noindent \textbf{The Sphere Covering Bound.} For any $q$, $n$ and $R$ the following holds:
	\begin{equation}
		K_{q}\left(n,R\right)\geq\frac{q^{n}}{\mathrm{\sum_{i=0}^{R}}{n \choose i}\left(q-1\right)^{i}}
	\end{equation}
	\\
	\indent As previously noted, by setting $n=\frac{q^{m}-1}{q-1}$ and $|W|=q^{n-m}$ we can obtain a perfect Hamming code with a covering radius of one, for which the sphere covering bound is attained with equality. For $n-1=\frac{q^{m}-1}{q-1}-1$ the sphere covering bound holds with strict inequality and thus,
	
	\begin{equation*}
		\begin{split}
		K_{q}\left(n-1,1\right) 
		&> \frac{q^{n-1}}{1+(n-1)(q-1)} \\[5pt]
		&= \frac{q^n}{q\left(1+(n-1)(q-1)\right)} \\[5pt]
		&> \frac{q^{n}}{q\left(1+n\left(q-1\right)\right)} = \frac{1}{q}K_{q}\left(n,1\right)
		\end{split}
	\end{equation*}
	\\
	\indent Therefore, we obtain that $K_{q}\left(n,1\right)<qK_{q}\left(n-1,1\right)$ for $n=\frac{q^{m}-1}{q-1}$. Thus, for any $1\leq k\leq n-1$ the following holds:
	
	\begin{equation*}
		K_{q}\left(n,1\right) \: < \: q^{k}K_{q}\left(n-k,1\right)
	\end{equation*}
	\\
	\indent This implies that $n=\frac{q^{m}-1}{q-1}$ is the minimal length required to obtain the seller's maximal value. \hfill $\qedsymbol$ \\
	
	Thus, not only is the lottery frame defined by the parameters of Hamming codes optimal, but it is also the minimal length required to maximize the seller's value. 
	
	Assuming that we have no restrictions on the alphabet size, the lottery's length and we allow the objective probability of winning to be arbitrarily small, the following corollary is immediate from the previous results.
	
	\begin{corollary}
		To maximize his value, the seller will choose $p=\frac{1}{q^{m}}$ where $q$ is a prime power, $m\to\infty$, $n=\frac{q^{m}-1}{q-1}$ and $|W|=q^{n-m}$.
	\end{corollary}

	Therefore, if the seller is not restricted in any sense he will opt for lotteries with a very small objective probability of winning that generate a relatively high near miss effect and with it a high perceived probability of winning. 

\section{Discussion}
	
This paper introduces a framework for analyzing the near miss effect in pure chance lotteries. We define a near miss index that quantifies the near miss effect a lottery induces and analyze the optimal lottery design. We show that a general construction of an optimal lottery frame that maximizes the near miss index, and with it, the perceived probability of winning is an open problem that generalizes the covering code problem. Nonetheless, our analysis and results show that clear comparisons between different lottery frames can be drawn, rendering specific frames preferable over others in terms of near miss.    

The analysis of the seller's optimal lottery design resulted in full characterization of the optimal lottery frame for that specific setting. We showed that the optimal lottery parameters belong to a special class of codes known as Hamming codes. Moreover, the results in this analysis might provide us with additional intuition and potential explanation as to why some of the most common lottery designs are those with a very small objective probability of winning.

Finally, The connection established in this paper between our near miss framework and the field of coding theory might open up new avenues of future research.

\newpage	
\nocite{*}
\bibliographystyle{aer}
\bibliography{document}

\end{document}